\documentclass[preprint2]{aastex}
\usepackage{graphics}

\newcommand{\chandra}{\hbox{\it CHANDRA}~}
\newcommand{\rosat}{\hbox{\it ROSAT}~}
\newcommand{\xmm}{\hbox{\it XMM--Newton}~}
\newcommand{\hi}{H{\sc i}~}
\newcommand{\hii}{H{\sc ii}~}

\newcommand{\halpha}{H{$\alpha$}~}

\newcommand{\kms}{$\rm km\,s^{-1}$ }
\newcommand{\pcms}{$\rm cm^{-2}$ }
\newcommand{\lumunit}{$\rm erg\,s^{-1} $}
\newcommand{\fluxunit}{$\rm erg\,cm^{-2}\,s^{-1}$}

\shorttitle{ROSAT observations of Holmberg II}
\shortauthors{Kerp, Walter \& Brinks}

\begin{document}


\title{ROSAT X-ray observations of the dwarf galaxy Holmberg II}


\author{J\"urgen Kerp}
\affil{Radioastronomisches Institut der Universit\"at Bonn,
              Auf dem H\"ugel 71, D-53121 Bonn, Germany}
\email{jkerp@astro.uni-bonn.de}

\author{Fabian Walter}
\affil{California Institute of Technology, Astronomy Department 105-24,
Pasadena, CA 91125, U.S.A.}
\email{fw@astro.caltech.edu}

\and

\author{Elias Brinks}
\affil{Departamento de Astronom\'{\i}a,
              Apartado Postal 144, Guanajuato, Gto. 36000, M\'exico}
\email{ebrinks@astro.ugto.mx}




\begin{abstract}
We present a study of the irregular dwarf galaxy
Holmberg~II based on  \rosat PSPC observations (total exposure time:
22\,ksec). Holmberg II is a nearby (3.2\,Mpc), well--studied dwarf
irregular galaxy. It is famous for its interstellar medium which is
dominated by expanding structures such as HI holes and shells. We search
for X--ray
emission from point sources as well as for diffuse emission, down to the
detection limit of the \rosat data. Using X--ray hardness
ratio diagrams we differentiate between
thermal plasma and power--law X--ray spectra which helps to determine
the nature of the individual sources. Correlating the
X--ray data with complementary observations ranging from the
far--ultraviolet to the radio regime we increase the probability of
correctly identifying sources belonging to Holmberg~II.  We did not
detect soft X--ray emission originating from hot gas within supergiant
\hi\ shells above our luminosity sensitivity limit of
($L_\mathrm{limit}\mathrm{(0.1 - 2.1\,keV)} \geq 10^{37}$\lumunit).
This finding can probably be attributed to blow--out in the case of the
largest holes
and insufficient sensitivity (due to strong photoelectric absorption) in
case of
the smaller \hi\ holes. However we find faint X--ray sources well beyond
the stellar
body but within the \hi\ distribution of Holmberg II, which suggests
the presence of X--ray binaries. This indicates that star
formation has taken place across the entire gaseous disk of Holmberg
II in the past, some of which may have created the structures seen in the
ISM
at large galactocentric radii.
\end{abstract}




\section{Introduction}

Holmberg II (hereafter abbreviated as Ho\,II) is one of the most
famous examples of a dwarf galaxy exhibiting a violent, disrupted ISM,
as traced by its appearance in the 21--cm \hi line \citep{puche}.
\hi\ holes are present across the entire galaxy ---
even well beyond Ho\,II's stellar body. \citet{puche}
compiled a catalog of individual \hi  shells. The measured
radial expansion velocities (10-25\,\kms) and diameters
(100--2000\,pc) indicate that most of the holes are still expanding and
have ages of $t \sim 10^7$ to $10^8$\,yr.

Ho\,II has been studied at many 
other wavelengths as well,  such as radio continuum 
wavelengths \citep{tonguewestpfahl}, optical \citep{rhode}
and FUV \citep{stewart}. Besides, it has been the
topic of studies to determine its mass distribution and dark matter
content \citep{bureau} and a popular galaxy
to compare numerical simulations/models of expanding \hi shells 
against \citep{mashchenko, oey}.

Recent \hi\ observations, including those of Ho\,II, show that \hi
holes are older and reach larger sizes in dwarf galaxies as compared
to similar structures in more massive, spiral galaxies. Some of this can
be
attributed to the fact that dwarf galaxies have shallower
gravitational potentials which leads, for the same observed \hi velocity
dispersion, to a puffed--up disk. Hence, \hi holes can grow to larger
sizes before breaking out of the disk. Moreover, dwarf galaxies
lack in general differential rotation or spiral density waves
(e.g., \citep{walterbrinks}). In other words, once formed,
structures like \hi\ holes in the ISM are not strongly modified and/or
destroyed by 
large scale streaming motions. 
                                                    
The most straightforward way to interpret the presence of the \hi
cavities is to attribute them to past stellar activity within their
centers (the 'standard picture'): strong winds of O-- and B--stars can
create individual bubbles of pc size whereas subsequent supernova
events create cavities of a few tens to hundreds of parsecs extent. In
this picture, the expansion of the \hi holes is powered by the
overpressure created within the coronal gas resulting from the SN
explosions, e.g. \citep{Wea77}.


Although the standard picture of the creation of the \hi\ holes is
widely accepted, it is not without its critics. For example \citet{rhode},
analyzed broad band BVR (4$\sigma$ detection
limit: B=23\,mag) and narrow band H$\alpha$ images to search for star
clusters in the centers of the \hi\ holes in Ho\,II. Based on the
results derived by \citet{puche} for the age and energy needed to
create the \hi\ hole, and assuming that the stellar cluster remained a
coherent structure during its evolution and using conventional initial
mass functions (IMFs), they argue that these star clusters should still
be detectable. As their observations revealed far less remnant
clusters than expected, \citet{rhode} ruled out
clusters as the origin for the \hi\ holes in all but 6 out of 48
cases. Certainly especially the giant holes outside the optical disk
of Ho\,II are difficult to explain since star formation currently
doesn't seem to play a dominating role at large galactocentric radii
of Ho\,II.

X--ray observations of dwarf galaxies can be used to test the validity of
the standard picture.
Coronal gas within the interiors of the \hi\ holes might be
detectable in X--rays, whereas objects related to the end points of
stellar
evolution -- such as X-ray binaries -- are expected to show up as hard
X-ray point sources.
Low mass X-ray binaries are prominent X-ray sources
detectable
even at the distance of Ho\,II. In the optical regime however, these
objects
are very  faint and difficult to detect even with large telescopes. 
X--ray observations seem the best way forward now to further check the
validity of the standard picture.

\citet{zezas} as well as \citet{miyaji} have reported \rosat data on the
strongest X--ray source in Ho\,II only.  Here, we perform an
analysis of all sources related to Ho\,II within the entire field
observed with the \rosat\ PSPC. Sec.\ 2 deals with the data
reduction and analysis of three pointed \rosat\ PSPC observations of
Ho\,II which we retrieved for that purpose. In Sec.\ 3 we introduce
the X--ray hardness ratio diagram as a tool to
classify the X--ray spectrum for faint sources. Our results are
presented in Sec.\ 4, which are then discussed in Sec.\ 5. We summarize our conclusions in Sec.\ 6.

\section{Data reduction and analysis}

\begin{deluxetable}{lrr}
\tabletypesize{\scriptsize}
\tablecaption{Sequence number of the ROSAT PSPC observations,
their date and integration time on source.\label{obstable}}
\tablewidth{0pt}
\tablehead{
\colhead{Seq. No.} & \colhead{date} & \colhead{$t_{\rm int }$}\\
\colhead{ } & \colhead{[mm/yy]} & \colhead{[s]}
}
\startdata
600140p & 04/92 & 7258\\
600431p & 10/92 & 11607\\
600431p-1 & 03/93 & 3701\\
\enddata
\end{deluxetable}

\subsection{Archival Data}
We retrieved 3 pointed PSPC observations towards Ho\,II from the
\rosat archive (see Table \ref{obstable} for details). The \rosat
PSPC data were analyzed using the EXSAS software package
provided by the ``Max--Planck--Institut f\"ur Extraterrestrische
Physik'' in Garching \citep{zimmermann}. All
three observations were merged into a single photon event file and
re--centered on the mean position: $\alpha_{2000} =
8^\mathrm{h} 19^\mathrm{m} 5^\mathrm{s}, \delta_{2000} = +70\degr
42\arcmin 36\arcsec$. The net integration time of the merged \rosat
PSPC data is 22566 sec, making Ho\,II one of deepest studied dwarf
galaxies by \rosat. The photon events were binned into the standard
\rosat 1/4 keV (also denoted as \rosat C-band), 3/4 keV (M-band),
and 1.5 keV (J-band) energy bands and a total energy band
(corresponding to the pulse--height invariant (PI) channels channels
11--41, 52--90, 91--201 and 11--201). We calculated a merged exposure
map for each individual X--ray image to overcome the problem of
radially decreasing sensitivity and vignetting of the \rosat PSPC.

The radius of the ROSAT PSPC point spread function (PSF) depends on
the X--ray photon energy and the off--axis angle with respect to the
optical
axis, resulting in a varying angular resolution between $23''$ and $34''$
within an off--axis radius of $10\arcmin$ from the optical axis.
We determined the X--ray
background (XRB) intensity level outside the \hi\ distribution of Ho\,II
($r
\geq 10\arcmin$), which we subsequently used as the ``off'' intensity
value.

We integrated the X--ray photons of each individual source within a
circular area with a diameter equal to 3.5 times the diameter of the
PSF ($3.5 \times \diameter_\mathrm{PSF}$) for each individual energy band. This area covers 99\% of the X--ray source photons.
From this total number of photons we subtracted the
contribution of the XRB, and evaluated the residual X--ray intensity
level in the immediate surrounding of the X--ray source. We classified
an X--ray source as detected if the number of net counts exceeded the
3$\sigma$ threshold above the noise set by the background level.

The results are shown in Fig.\ \ref{fig:hi-xray} top panel which shows the
\rosat\ PSPC 1/4\,keV map as contours superimposed on a
grey-scale rendering of the neutral hydrogen distribution 
(\citet{puche}). The bottom panel shows the X--ray
intensity distribution averaged over the entire 0.1--2.1\,keV energy
range.  Fig.\ \ref{fig:complementary} displays this latter map in the top
panel as contours superimposed on a 4.86\,GHz radio continuum image 
\citep{tonguewestpfahl} whereas the contours in the
bottom panel are overlayed on a map of the \halpha\ emission.

\subsection{Confusion with cosmic X--ray background sources}

The extragalactic XRB is due to the superposed emission of individual
objects, mostly active galactic nuclei (AGNs) and quasi stellar
objects (QSOs) \citep{hasinger,mushotzky}.
Unfortunately, the angular resolution of the
\rosat PSPC does not allow one to resolve faint individual XRB sources,
i.e., the PSPC is confusion limited. To constrain the contribution of the
diffuse extragalactic XRB radiation, one has to evaluate the overall
spectral characteristics of this X--ray emission component.  
\citet{gendreau} showed that the extragalactic XRB emission
can be approximated by a power--law with $I(E) \propto E^{-\Gamma}$
and $\Gamma \simeq 1.5$. They determined an intensity of the
extragalactic XRB at 1 keV of about 9.6
$\mathrm{keV\,cm^{-2}\,s^{-1}\,sr^{-1}\,keV^{-1}}$.

Taking this value for the XRB and assuming that the extragalactic XRB
is only attenuated by the interstellar medium (ISM) of the Milky Way,
we derive within the area of an individual X--ray source towards Ho\,II
($3.5
\times \diameter_\mathrm{PSF}$), an X--ray flux for the extragalactic
XRB of about $F_\mathrm{XRB} = 2.8 \times 10^{-15}$\fluxunit.

So far, we have been focusing on the overall extragalactic XRB flux
level.  However, we have to take into account the log($N$)
vs.\ log($S$) relation \citep{hasinger} to evaluate
the frequency of extragalactic XRB sources at a certain flux level.
Unfortunately, there is no straightforward way to discriminate between
unrelated background sources and those belonging to Ho\,II.  However,
we can reduce this ambiguity by studying the X--ray colors of a source
and by searching for counterparts at other wavelengths using
complementary data (which is what we will do in the following).
Extragalactic sources unrelated to Ho\,II -- and absorbed
by its gas distribution -- will appear as hard X-ray sources,
detectable preferentially in the \rosat M-- and J--band. 
The 3$\sigma$ X--ray flux level across the entire \rosat energy band
is $F_\mathrm{3\sigma}{\rm (0.1 - 2.1\,keV)} = (4.1\pm1.4) \times
10^{-15}$\fluxunit, above the diffuse extragalactic background level.
Applying the  log($N$) vs.\ log($S$) relation derived by
\citet{hasinger} to the
area of Ho\,II \hi gas distribution, we expect to detect about 7
significant
X-ray sources unrelated to Ho\,II.

Assuming a distance to this object of 3.2 Mpc
and a galactic X--ray attenuating column density of 
$N_{\rm HI} = \sim 3 \times 10^{20}$ \pcms \citep{hartmann}, we
derive a detection luminosity threshold of $L_{\rm 3\sigma} \geq (10.8
\pm 3.6) \times 10^{36}$\lumunit for sources associated with Ho\,II
(neglecting for the moment absorption within this galaxy). Only very young
supernova remnants 
or accreting X--ray binaries typically exceed this threshold.
The \rosat PSPC X--ray data presented here therefore trace
only two extreme populations in Ho\,II: the actual ($t < 10^5$ yr,
i.e., supernova remnants) or a population older than $t > 10^7$\,yr,
such as pulsars or accreting X--ray binaries. The intermediate age
stellar population can be traced via \halpha\ and far-ultraviolet
(FUV) observations \citep{stewart}. Combining all
data provides a unique picture of the star formation history of Ho\,II
over the last 100 Myr.

\section{The Hardness Ratio Diagram}

In this section we will introduce a useful tool to constrain the spectral 
properties of faint X-ray sources.
Using this tool it becomes possible to estimate whether the spectrum of a
(faint)
X--ray source is predominantly characterized by a thermal emission
spectrum or
rather by a power--law. We
will first discuss briefly the characteristics of those sources likely
to be encountered in our object and then proceed to explain what we
call the Hardness Ratio Diagram and where these
different sources are likely to be encountered in these diagnostic
diagrams. 

Supernova remnants are characterized by their thermal plasma
radiation. Below a temperature of $T\simeq 10^{6.4}$\,K, emission
lines dominate the X--ray spectrum. Above this temperature,
Bremsstrahlung is the most important cooling mechanism of the hot
plasma. The cooling time of the plasma can be approximated via $t_{\rm
cool} \simeq \frac{3}{2} \frac{N {\rm k} T}{\Lambda_{\rm N} n_{\rm t}
n_{\rm e}} \propto ~\frac{\sqrt{T}}{n_{\rm e}}$ where $N$ is the total
number of 
particles,
$T$ is the hot plasma temperature in [K], ${n_{\rm t}}$ the ion density,
${n_{\rm e}}$ the electron density, both in [cm$^{-3}$] and $\Lambda_{\rm N}$
the normalized
cooling rate \citep{sutherlanddopita}. This leads to $t_{\rm cool} \simeq
10^9$\,yr,
assuming a temperature of $T\simeq 10^{6.4}$\,K and a typical electron
density of about $n_{\rm e} = 2 \times 10^{-3}\,{\rm cm^{-3}}$.
The expected X--ray luminosity of a single young SN is 
$L_{\rm X-ray} \sim 10^{38}$ \lumunit \citep{fabbiano96}.

Core--collapse SN are correlated in space and time, their progenitor
having been formed in groups, OB associations or even larger
conglomerates. Expanding individual supernova remnants are therefore
likely to merge, creating expanding shells referred to in the
literature as super-bubbles or supergiant shells which are supposedly
filled with tenuous, high temperature gas. These shells show up in
\hi\ maps as holes in the neutral hydrogen distribution.
In the case of Ho\,II the largest holes have diameters roughly twice
the $1\sigma$ scale height of the \hi\ disk of 600\,pc, according to
\citet{puche}. The cooling time of the coronal gas within the interior of
the shells is much longer than the ages of the \hi\ 
holes which should make them detectable in X--rays. However, the
largest super-shells, with diameters exceeding
1\,kpc, likely broke  out of the disk, at which moment
the hot gas interior is vented in to the halo, the hot plasma is lost
and the hole becomes invisible in X--rays. In the case of the smaller,
still confined holes, the interior X--ray emission should be
detectable if the \hi\ column density of the approaching side of the \hi\
shell
does not exceed values of $N_{\rm HI} \geq 5 \times 10^{20}$\,cm$^{-2}$,
above which
value the soft X--ray emission ($E \leq 0.5$ keV) of the hot gas is
absorbed {\it in situ}
by the shell.

Pulsars, cooling neutron stars as well as accreting X--ray binaries
are the very end points of stellar evolution and may be observable
within the interior of the holes, under the assumption that the origin
of the holes is caused by star forming activity and subsequent rapid
stellar evolution of the most massive stars. Their X--ray luminosities
typically range between $L_\mathrm{X}(\mathrm{0.1 - 2.4 keV}) \simeq
10^{36}$--$10^{38} {\rm erg\,s^{-1}}$.

To identify the emission mechanism of an X--ray source the standard
procedure is to extract a spectrum from the X--ray data. The
brightest source associated with Ho\,II has previously been studied in this way
by \citet{zezas}. However, due to the large
number of free parameters, even the analysis of this high
signal--to--noise X--ray spectrum is not without its ambiguities.

In the case of faint X--ray sources ($F_{\mathrm X} \simeq 5 - 20
\times 10^{-15}$ \fluxunit), which are discussed here, the
insufficient signal--to--noise ratio does not allow us to extract
much information from the \rosat PSPC spectra.  However, the
count rates within the individual broad \rosat energy bands contain
significant
information about the X--ray source spectrum. We therefore decided to
study the hardness ratios to constrain the X--ray emission process.
For this aim, we subtracted the contribution of the extragalactic X-ray
background emission from the count rates of the individual X-ray sources.
In Fig.\ \ref{colourcolour} we plot the \rosat hardness ratio 1 (HR1)
($\frac{{\rm M+J-C}}{{\rm C+M+J}}$) versus the \rosat hardness ratio
2 (HR2) ($\frac{{\rm J-M}}{{\rm M+J}}$). The stronger the
photoelectric absorption the harder the resulting X--ray spectrum. The
lines in the figure indicate the locus where one can expect individual
sources to fall in the hardness ratio diagram,
depending on the type of X--ray emission, either thermal or
power--law, over a range of temperatures and power--law indices,
respectively, and over a range of foreground absorptions. This latter
effect is quite comparable to the reddening caused by intervening dust
in an optical color--color diagram.

The relations displayed in Fig.\ \ref{colourcolour}, both for the
power--law (solid lines) as well as the thermal plasma spectra -- assuming
solar metal abundances --  (dashed
lines), are plotted for increasing absorbing foreground column density
(from left to right $N_\mathrm{HI} = 0, 2, 3,$ and $ 4 \times
10^{20}\,\mathrm{cm^{-2}}$).  The leftmost relation of both the
power--law or the thermal case represents the un-absorbed
situation. The numbers along the tracks for $N_\mathrm{HI} = 3
\times 10^{20}\mathrm{cm^{-2}}$ (which corresponds to the absorbing
foreground column density corresponding to the Milky Way;
\citep{hartmann} give the spectral slope for the power--law
spectra (solid lines) and log($T$[K]) in case of the thermal plasma
spectra. Obviously, soft X--ray sources will predominantly be
found in the lower left part of the X--ray hardness ratio
diagram, whereas highly absorbed or intrinsically
hard X--ray sources populate the upper right part of the
diagram.

\section{Results}


\begin{figure*}
\rotatebox{-90}{\resizebox{6cm}{!}{\includegraphics{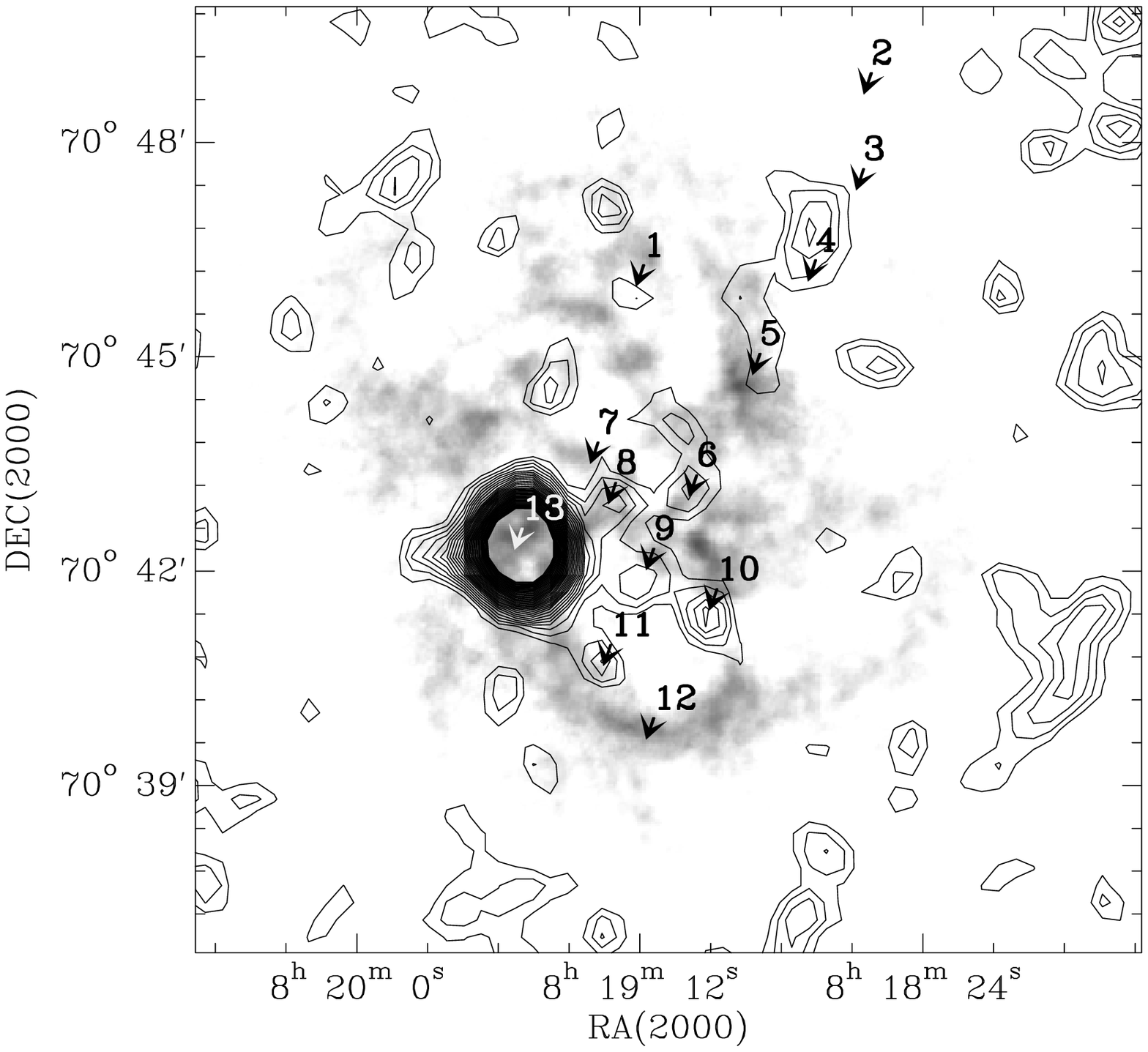}}}
\rotatebox{-90}{\resizebox{6cm}{!}{\includegraphics{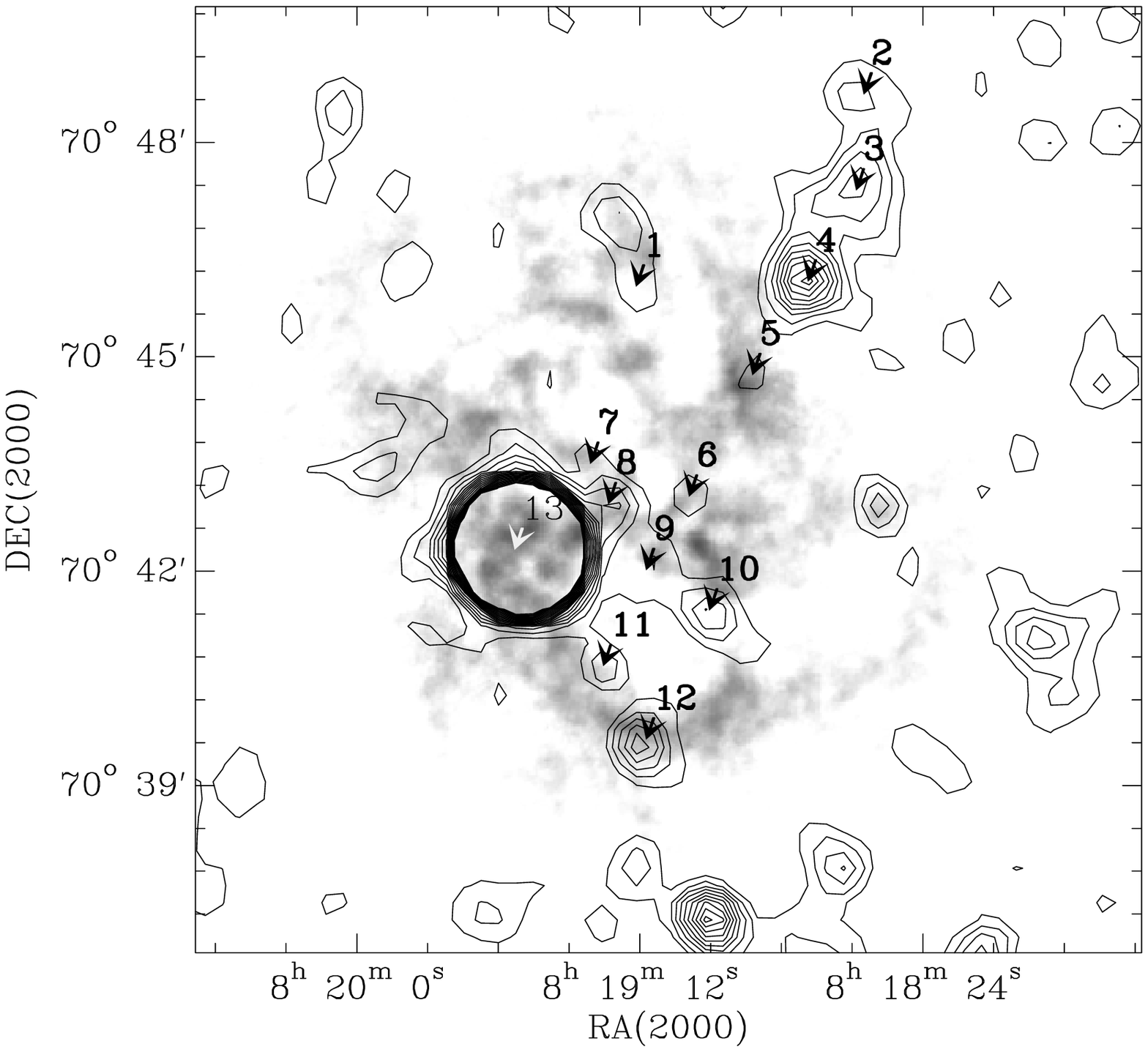}}}
\figcaption{{\bf Left:} \rosat PSPC 1/4 keV intensity
distribution superimposed as contours on the \hi\ 21-cm map of
Ho\,II. The numbers denote the X--ray sources according to
Table \ref{xraysources}. The arrows point to the center of
each X--ray source. The contour lines start at the 3$\sigma$ level,
incrementing in steps of 1$\sigma$ (angular resolution smoothed to
75"). {\bf Right:} \rosat PSPC total
(0.1 keV -- 2.1 keV) X--ray intensity distribution superimposed as
contours on the \hi 21-cm map of Ho\,II.  The contour lines start at
the 3$\sigma$ level incrementing in steps of 2$\sigma$.  Note that the
individual sources \#2, \#3, \#4 and \#12 are not positionally
coincident with 1/4 keV X--ray sources and thus may be unrelated
background sources (see discussion in the text).
\label{fig:hi-xray}}
\end{figure*}

In total we detected 31 significant X--ray sources within the extent
of the \hi\ distribution of Ho\,II. As mentioned earlier,
only 7 X-ray sources are expected, based on the log($N$) vs. log($S$) relation
derived by \citet{hasinger}.
To avoid any confusion with unrelated extragalactic XRB sources, we
decided to study only those X--ray sources which are identified in at
least one additional frequency range. This
complementary information also gives us a better handle on the nature
of the X--ray source in question. This left us with 13 sources, all
believed to be related to Ho\,II. The properties of these 13 sources
(positions, fluxes, luminosities, complementary data) are listed in
Table \ref{xraysources}. We mark those sources with an ``x'' which
emit significant soft (1/4 keV) X--ray photons, implying that they are
most likely associated with Ho\,II.
Because a background X-ray source will be attenuated by the entire
amount of ISM 
belonging to Ho\,II, this leads to strong photoelectric absorption,
reducing especially the soft part of the X-ray spectrum.
We tried to constrain the X--ray
emission process further by studying the spectral properties of the
X--ray sources via the hardness ratio diagram (Sec.~3).

\subsection{X--ray sources inside \hi\ holes}

The \rosat 1/4 keV map (Fig.\ \ref{fig:hi-xray}) shows the emission
which is most likely associated with X--ray plasma radiation. Because
of the fairly high luminosity limit of the \rosat observations
($L_\mathrm{limit}\mathrm{(0.1 - 2.1\,keV)} \geq 10^{37}$\lumunit),
supernova remnants are the most likely candidates to be encountered.

\begin{figure*}
\rotatebox{-90}{\resizebox{6cm}{!}{\includegraphics{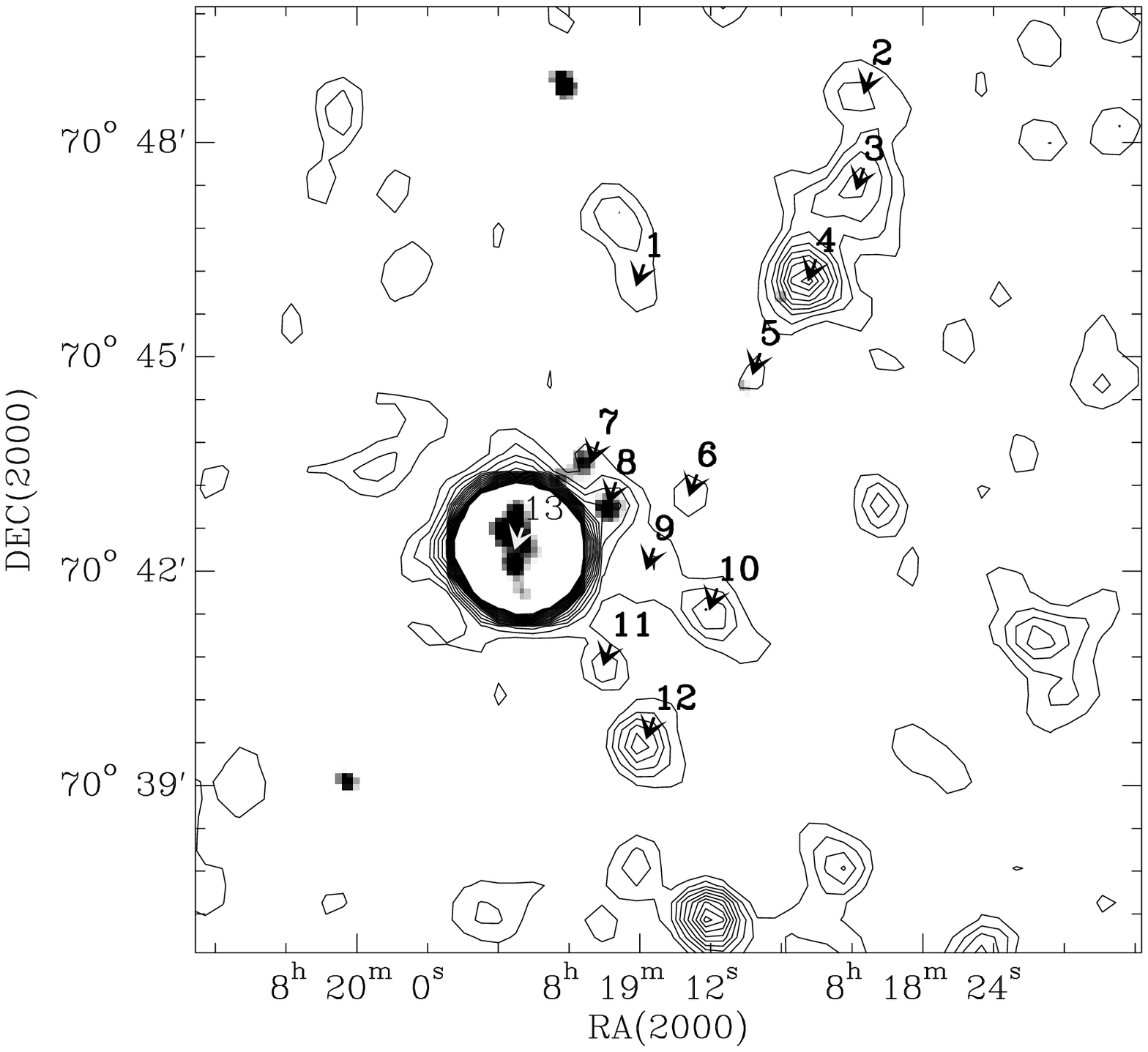}}}
\rotatebox{-90}{\resizebox{6cm}{!}{\includegraphics{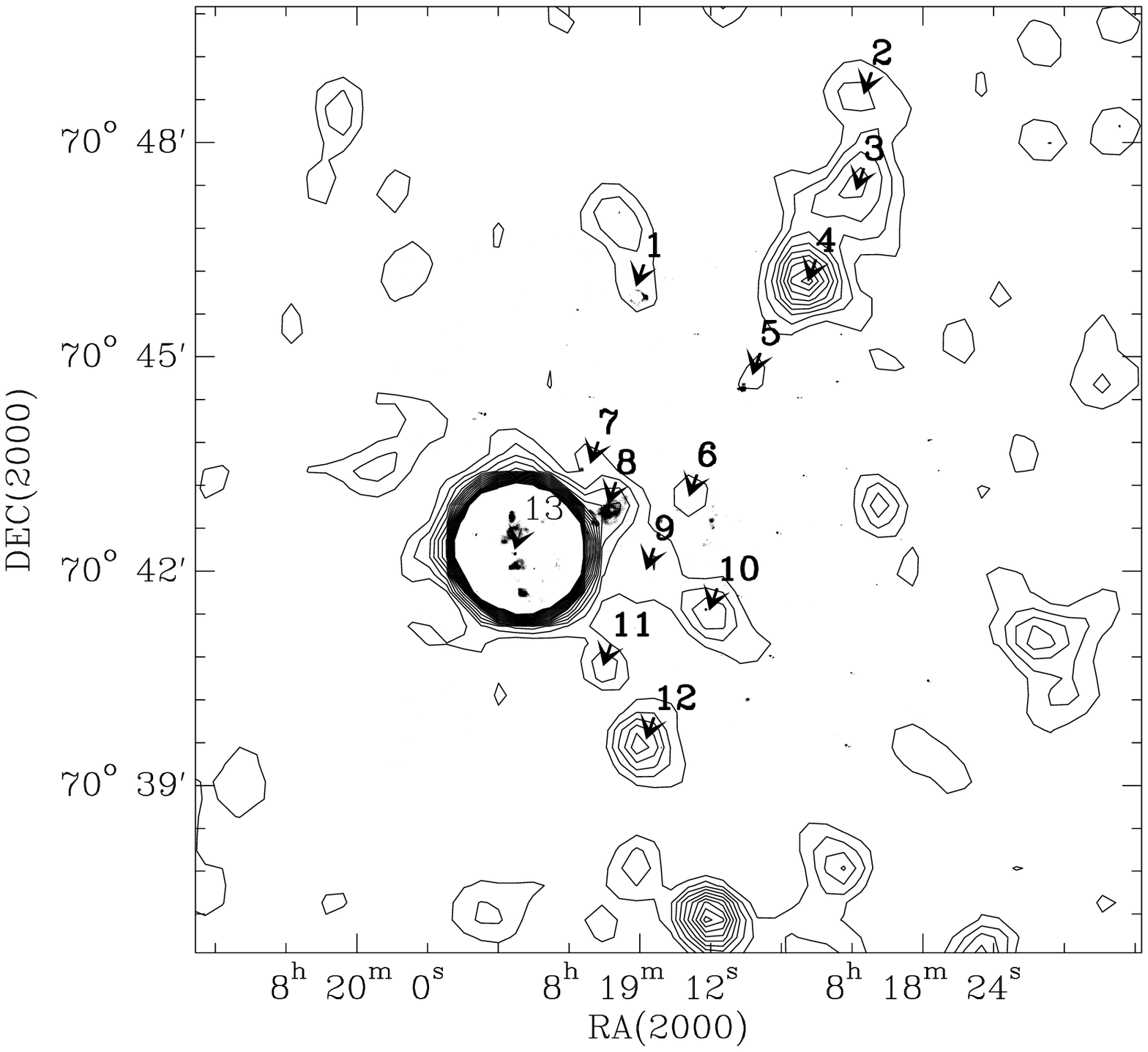}}}
\figcaption{{\bf Left:} Superimposed on a gray--scale
representation of the 4.86 GHz radio continuum VLA map of Ho\,II are
the contour lines of the total (0.1 keV -- 2.1 keV) X--ray intensity
distribution. The contours start at 3$\sigma$ incrementing in steps of
2$\sigma$ ( angular resolution smoothed to 75").
{\bf Right:} \halpha\ map of Ho\,II superimposed by the
contour lines corresponding to the total \rosat energy window, as
presented om the left.  The available \halpha\ map only covers  parts of the
region of interest. \label{fig:complementary}}
\end{figure*}

The X--ray sources \#6, \#7, \#9, \#10 and \#13 are associated with
\hi\ holes cataloged by \citet{puche} their centers
are located within the extent of
the \hi\ holes (for details see Table \ref{xraysources}). Although
these soft X--ray sources are located within the low volume density
cavities of Ho\,II, one cannot exclude a chance association with an
extragalactic X--ray background source. The X--ray sources which most
probably belong to Ho\,II should be detectable simultaneously in the
radio and/or the optical regime. Sources \#7, \#8, and \#13 are
positionally coincident with radio continuum emission (see Fig.\
\ref{fig:complementary}(top) and also \citet{tonguewestpfahl}).

According to \citet{tonguewestpfahl} the X--ray
source \#7 (\hi\ hole No. 36) coincides with a steep spectrum
non--thermal radio source. In Fig.\ \ref{diagramme} we plot all X--ray
sources brighter than $11\times 10^{-15}$ \fluxunit\ in a hardness ratio
diagram similar to the one presented in Fig.\
\ref{colourcolour}. This suggests for source \#7 a steep X--ray
energy power--law spectrum ($I(E) \propto E^{-2.5}$), implying an
accreting X--ray binary system as a likely source of the radiation
\citep{white}. \citet{tonguewestpfahl} favor an extragalactic origin,
unrelated to
Ho\,II, because they interpret the steep radio continuum spectrum as
an indication for spectral aging of the electron population which is
frequently discussed in the framework of evolutionary models of radio galaxies.
However, the
X--ray as well as the positionally coincident radio continuum source
fits very well within the \hi\ hole No.\ 36 of the \citet{puche}
catalog. Moreover, close to X--ray source \#7 a 
small \hii region is located (Fig. \ref{fig:complementary}). One may
speculate that this \hii region and associated \halpha\ emission is
due to secondary star formation, on the rim of \hi\ hole No.\  36 (Puche
et al. 1992).

The X--ray sources \#6, \#9 and \#10 are not exactly coincident with
\halpha\ or radio continuum emission. However, \halpha\ and radio
continuum radiation is present within the $3.5 \times
\diameter_\mathrm{PSF}$ diameter discussed earlier which is why we
included these sources in our analysis. Close to the southern boundary
of \#6 (\hi\ holes No. 22 and No. 23), \halpha\ and thermal radio
continuum emission indicates the presence of a patchy star forming region. The FUV
data of \citet{stewart} suggest a young star forming
region with an age of 2.5 -- 3.5 Myr. Inspection of Table~2 of
\citet{rhode} supports the view that X--ray source \#6
(positionally coincident with \hi\ hole No. 22) is associated with a
stellar cluster.

The X--ray sources \#9 (\hi\ hole No. 29 and No. 32) and \#10 (\hi\
hole No.  21) are associated with faint \halpha\ and FUV emission at
their rims.  The FUV data implies an age of $ > 6.3$\,Myr
\citep{stewart}. This is
consistent with the findings based on the X--ray hardness ratio
diagram (Fig.\ \ref{diagramme}) and suggests
power--law spectra for both X--ray emitting sources.
Their X--ray luminosities of a few $10^{37}$\lumunit\
classify both as likely candidates of accreting binary systems
\citep{white}.

\subsection{X--ray sources outside \hi\ holes}

Sources \#1, \#5, \#8 and \#11 are located outside \hi\ holes,
towards high column density regions of Ho\,II. The accidental
association with an extragalactic XRB source becomes much less likely,
because of the strong attenuation of a possible unrelated extragalactic
soft
X--ray source by the significant amounts of X--ray absorbing matter
within Ho\,II (in addition to the roughly constant absorption due to
the Milky Way).

Source \#1 is associated with bright \halpha\ and FUV emission,
however it is not detected in radio continuum maps. This can be
interpreted as an indication for a developed star cluster
with evolved stars, which already partly left the main sequence.
The FUV data implies an age of $ > 6.3$\,Myr. The X--ray emission is
too faint to set constraints on the most likely X--ray emission process,
even when employing the hardness ratio diagram. So, no
firm
conclusions as to the nature of the source can be derived from the
available data.

Source \#5, which shows X--ray emission at the 3$\sigma$ level,
coincides with a radio continuum source.
\citet{tonguewestpfahl} derived a thermal radio continuum
spectrum for that source from their multi--frequency analysis.  This
is consistent with the \halpha\ map (bottom panel of Fig.\
\ref{fig:complementary}) which shows a compact \hii region at this
position. Therefore the most likely conclusion regarding the nature of
this source is that of a young star forming region.

Source \#8 emits
thermal radio continuum emission as well. This is consistent with the
\halpha\ map of Ho\,II (Fig.\ \ref{fig:complementary}) which reveals a
very prominent \hii region at this location. The linear extent of
the \hii region is of order 100 pc. The X--ray hardness
ratio diagram also implies a thermal origin of the
detected X--ray emission (log($T$[K]) $\simeq$ 7).  A single luminous
O star has a luminosity of about $L_\mathrm{X-ray}(\mathrm{O-star)}
\simeq 10^{34}$\lumunit\ in the \rosat X--ray energy window
\citep{kudritzki}. Therefore, we would require a
contribution by several hundred young stars in order to account for
the observed X--ray emission. Because the radio continuum spectrum is
thermal in nature, the region has to be very young as otherwise
supernovae and their remnants would start to dominate the radio
continuum spectrum with a non-thermal component.

Taking into consideration that the \halpha\ ridge related to source
\#8 is located between the boundaries of two \hi\ holes (No.  29 and
No. 36), very recent induced star formation may be taking place here.
\citet{stewart} derived an age of 3.5 -- 4.5 Myr for
that particular region, consistent with the presence of a young
massive star forming region. Additional evidence for a stellar cluster
is provided by the optical search for stellar emission within the \hi\
holes performed by \citet{rhode} who report that a
stellar cluster may underlie the nearby \hi\ hole No. 36.

X--ray source \#11 is of interest because the analysis of
\citet{stewart} implies a young star forming region,
with an age of 2.5 -- 3.5 Myr, whereas this region is not detected in
the \halpha\ map. Its location in the X--ray hardness ratio
diagram implies a power--law X--ray spectrum, indicating an accreting
compact object. Both the absence of \halpha\ radiation and the X--ray
data suggest that towards this high column density region we observe
more likely the end point of stellar evolution, like an X--ray binary.

\subsection{X--ray sources outside the stellar disk}

Sources \#2, \#3, \#4 and \#12 are located far beyond the
stellar distribution. Sources \#2 and \#4 are associated with faint
radio continuum emission. Both sources are prominent within the
\rosat 3/4 keV and the 1.5 keV energy band, whereas in the 1/4 keV
map they are only seen at the $3\sigma$ level. This indicates that
both sources have a hard intrinsic X--ray spectrum or, alternatively,
are highly absorbed X--ray sources. The hardness ratio diagram implies in the case of source \#2, that its emission may be
associated with thermal plasma radiation. In the case of source \#4
it is not possible to differentiate between a power--law and a thermal
plasma X--ray spectrum.

Source \#3 is located in between both sources. These three sources
appear in the X--ray map as a coherent structure. Source \#3 is
visible at high signal-to-noise level in the \rosat 1/4 keV map
(Fig.\ \ref{fig:hi-xray}). Together with the fact that it is situated
towards high \hi\ column densities suggests that this source is
associated with Ho\,II and not located beyond its $N_{\rm HI}$ column 
density distribution. This is the reason why we decided to include
this source in Table \ref{xraysources}. However, the complementary
data do not reveal any obvious counterpart for source \#3.

Finally, source \#12 is located towards a high \hi\ column density
region of Ho\,II. Neither in the radio continuum nor in the \halpha\
map (Fig.\ \ref{fig:complementary}) is there any emission associated
with this particular source. The FUV map of \citet{stewart}, however,
revealed excess emission positionally
coincident with this particular source. Quite comparable to the case
of source \#11 (see discussion above) the analysis of
\citet{stewart} argues for a young star forming region of 2.5 -- 3.5
Myr, whereas the X--ray data does not exclude the possibility that an
end point of stellar evolution may account for the observed X--ray
emission. This hypothesis is furthermore supported by the absence of
\halpha\ emission. The X--ray luminosity of source \#12, with
$L_\mathrm{\#12} = (2.28\, \pm\, 0.32) \times 10^{38}$ \lumunit, is
compatible with both a supernova remnant or an accreting binary
system.

\subsection{The brightest X--ray source in Ho\,II}

The brightest X--ray source, labeled \#13 in this paper, has
been the subject of papers by \citet{zezas} and \citet{miyaji}.
This source dominates the entire X-ray emission of Ho\,II, because 87\% of all detected photons emerge from this unique object. Without its contribution the total X--ray luminosity of Ho\,II would only be about $L_{\rm X} \simeq 1\times 10^{39}\,{\rm erg}$, which is equivalent to the added X-ray luminosities of about six high mass X-ray binaries. In this respect Ho\,II is not a spectacular object in X-rays.
Here we reanalyze this object using the X--ray hardness ratio
diagram. This source can be very well
approximated by a single--component thermal X--ray spectrum with a
temperature of log($T$[K]) = 7.0 and an absorbing column density of
$N_{\rm HI} = 4 \times 10^{20}\,\mathrm{cm^{-2}}$ (the dashed line
to the very right). The neutral hydrogen column density of
$N_{\rm HI} = 3 \times 10^{20}\,\mathrm{cm^{-2}}$ (dashed line
with the log($T$[K]) values) does not fit the hardness ratios of
source \#13 within the uncertainties.  This implies that source \#13
is located within a low density cavity within the interstellar medium
of Ho\,II, as at least $3 \times 10^{20}\,\mathrm{cm^{-2}}$ is due to
Galactic foreground extinction, leaving at most $N_{\rm HI} =
1 \times 10^{20}\,\mathrm{cm^{-2}}$ to any residual column density of gas
belonging to the gaseous body of Ho\,II along the line of sight to
source \#13. Comparing the plasma temperature based on the X--ray
hardness ratio diagram with the results of
\citet{zezas},
our temperature value falls in between the
extreme temperatures derived by them.  The X--ray attenuating column
density implied by the X--ray hardness ratio diagram, however, is
significantly lower than that derived by these authors.

\begin{figure}
\epsscale{0.7}
\rotatebox{-90}{
\plotone{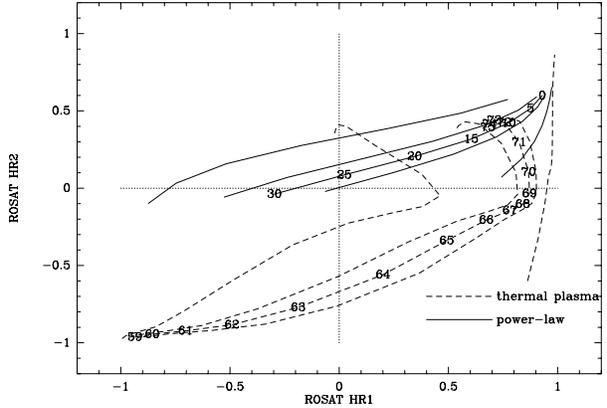}
}
\caption{Hardness ratio diagram. The solid lines
represent the expected hardness ratio locus for power--law X--ray
spectra; the dashed lines mark thermal plasma spectra.  The tracks
have as free parameter the X--ray absorbing column density.  Starting
left--most with the un-absorbed situation ($N_{\rm HI} = 0, 2, 3,$
and $4 \times 10^{20}\mathrm{cm^{-2}}$).  The numbers along the
$N_{\rm HI} = 3 \times 10^{20}\mathrm{cm^{-2}}$ track give the
energy index in case of power--law and log($T$[K]) in the case of
thermal plasma spectra.
To illustrate the case of very high column densities, we also plotted
the color--color tracks for $N_{\rm HI} = 10\times
10^{20}\mathrm{cm^{-2}}$, both for thermal plasma emission 
as well as for power-law X-ray spectra.
\label{colourcolour}}
\end{figure}

If one wanted to explain the hardness ratios of source \#13 with a
power--law spectrum one would have to assume that the X--ray spectrum
of this source is highly absorbed (see Fig. \ref{colourcolour}).
None of the four plotted
tracks representing the power--law X--ray spectra fit the
position of source \#13 in the hardness ratio diagram. Inspecting Table~2 of \citet{zezas} shows
that a power--law X--ray spectrum would require an absorbing column
density
of about $N_{\rm HI} = 1.2 \times 10^{21}\,\mathrm{cm^{-2}}$,
implying that according to the $N_{\rm HI}$ of
\citet{puche}, source \#13 would have to be located on the far side
of Ho\,II or deeply embedded within a molecular cloud.

To constrain the nature of source \#13 further, we consider the X--ray
luminosities of likely X--ray emitting candidates. Accreting binaries
have typical luminosities of $L_{\mathrm X} \simeq
10^{36}\,-\,10^{38}$\lumunit. Source \#13 however has $L_{\mathrm X} \geq
10^{40}$\lumunit. On the other hand, several supernova events are
observed with reported X--ray luminosities of $L_{\mathrm X} \geq
10^{40}$\lumunit\ \citep{fabbiano, fabianterlevich}.
The supernova hypothesis appears
to be a more likely explanation than a highly unusual X--ray binary
system, which exceeds the Eddington limit.
The supernova hypothesis is further supported by the radio
continuum data \citep{tonguewestpfahl}.  They
classified a positionally coincident radio continuum source as a
supernova remnant.

Partial support for this interpretation comes from \citet{miyaji} who
applied a wobble correction to the \rosat\ HRI image of Ho\,II and
claim that 25\% of the emission from this source is due to an
extended component with a diameter of order $10''$. Their {\it ASCA}
data reveal, in addition to confirming a soft component which they
attribute to (multiple) supernovae, a power--law component
accounting for most of the flux density beyond 2.5 keV, implying the
presence of an intermediate--mass black hole. A more definitive
description of the nature of source \#13 will have to await \chandra
or {\it XMM--Newton}. 

\begin{figure}
\epsscale{0.7}
\rotatebox{-90}{
\plotone{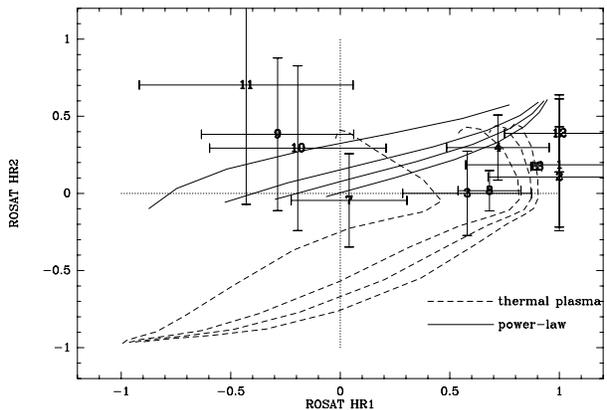}
}
\caption{Hardness ratio diagram as in Fig.\
\ref{colourcolour}. The X--ray sources brighter than
$F_\mathrm{X} \geq \times 10^{-14}$\fluxunit are
labeled according to the numbers given in Table \ref{xraysources}.
\label{diagramme}}
\end{figure}

\section{Discussion}
\subsection{Unresolved Sources}
A casual inspection of Fig.\ \ref{diagramme} shows that most of
the X--ray sources of interest lie above the track with
$N_{\mathrm{HI}} = 3 \times 10^{20}\,\mathrm{cm^{-2}}$ which is in
excellent agreement with what is expected from foreground extinction due
to the Milky Way \citep{hartmann}. This is an
important result as it implies that the X--ray sources are not deeply
embedded within the interstellar medium of Ho\,II.

Of the five X--ray sources located within the interiors of \hi\ holes
cataloged by \citet{puche}, two of them appear to be
associated with supernova remnants (\#7 and \#13), one can be
identified with a young star-forming region (\#6) and the remaining two
are most likely associated with the end points of stellar evolution,
more specifically accreting X--ray binaries (\#9 and \#10).

Four X--ray sources are located towards high column density regions of
Ho\,II. Because of their location in the soft part of the X-ray hardness ratio 
diagram, and given that they would have been fully absorbed 
if they were located beyond the high column density regions, they must be
in front of most of the \hi\ and hence it is likely that they are
sources related to Ho\,II. One may be associated with an evolved
stellar cluster (\#1), two
with young star forming regions (\#5 and \#8) and one with an
accreting X--ray binary system (\#11).

We find four X--ray sources which are located far beyond the stellar
body and close to the edges of the \hi\ column density distribution of
Ho\,II. All sources are either supernova remnants or X--ray
binaries. The importance here is that even though there is no evidence
based on optical imaging \citep{rhode} of past star
formation, this implies that star formation has occurred some
$10^8$\,yr ago at large galactocentric radii. This suggests that 
in general the supergiant \hi\ holes may
indeed be residuals from energetic events associated with past star
forming activity of Ho\,II, on timescales $t > 10^9$\,yr.

As most sources are too weak to derive an X--ray spectrum, we developed
a classification based solely on the hardness ratio diagram. The faintest source analyzed in this way has an X--ray flux
of only $F_\mathrm{X}\mathrm{(0.1-2.4\,keV)}\,=\,11\times 10^{-15}$
\fluxunit. Although not conclusive, the hardness ratio diagram
allows a separation between sources with thermal and power--law
spectra.

To reach a firmer conclusion as to the nature of the X--ray sources we
combined all available information ranging from the X--ray to the
low--frequency radio regime. All but one of the X--ray
sources, \#2, are associated with FUV emission, as presented by
\citet{stewart}, confirming their spatial correlation
with regions of recent, massive star formation.

Some of the X--ray emission of Ho\,II is associated with \hii regions.
This correlation remains somewhat tentative, however, as we 
have to take into account that the X--ray detection threshold is high,
strongly selecting only the brightest star forming regions. More
sensitive \chandra\ and \xmm\ observations will overcome this
instrumental limitation.  In the case of X--ray source \#8, the
H$\alpha$, radio continuum and X--ray data indicate independently the
presence of a massive star forming region.  Because of its location in
between two prominent \hi\ holes, we speculate that one of the holes
might be re-heated by a recent supernova event (X--ray source \#7);
one may then further speculate that this is evidence for induced star
formation.

\subsection{Extended Emission}

The  \rosat observations reported here did not reveal any hot
gas associated with the \hi\ holes. At first sight this might seem
remarkable. For example, in the Milky Way several hot--plasma filled
HI cavities are known.  The most prominent one is the local
bubble which encloses the Sun \citep{sfeir}. Its
X--ray emission is very soft (log($T$[K]) $\leq$ 6.0) and
faint. However, its pressure of $P_\mathrm{local} \simeq 10^4\,
\mathrm{K\,cm^{-3}}$ is a factor of 2 to 5 higher than that of the
neutral clouds located within the cavity. Hence, this local cavity
presumably is a low volume density region produced by several
supernova events, the latest supernova event dating back to
$1 - 2 \times 10^6$ years. Another interesting Galactic feature is the X--ray
bright super-shell known as Loop I or North Polar Spur
\citep{eggeraschenbach}.  Loop I has a diameter of about
$\sim$ 300\,pc. Its temperature is log($T_\mathrm{Loop I}$[K]) = 6.7,
much higher than that of the local X--ray plasma.  It is highly
over pressured ($P_\mathrm{Loop\,I}\,=\,2.5 \times 10^4\,\mathrm{K\,
cm^{-3}}$) and still expanding.

Observed  with \rosat from a hypothetical vantage point, such as an
external galaxy, both X--ray features 
would have remained undetected because the photoelectric absorption of
the enclosing thin neutral shell with $N_\mathrm{HI}\,\sim\,3 \times
10^{20}\mathrm{cm^{-2}}$ is sufficiently high for the X--ray photons
originating within the interior of the shell to be absorbed.  The only
extragalactic detections of hot gas in super-bubbles claimed thus far
are restricted to dwarf galaxies. Examples are the supergiant shell
LMC\,4 \citep{Bom94}, and the super-bubbles N\,44 and N\,11
\citep{Chu93,Mac98}, all three
situated in the LMC. Other examples are the supergiant shell SGS\,2 in
NGC\,4449 \citep{Bom97} and the possible super-shell
near Holmberg~IX \citep{Mil95}.

Recently \citet{Wal98} detected marginally extended
X--ray emission coinciding with a supergiant shell in IC\,2574
in a pointed \rosat observation. They found an
X--ray luminosity ($D$ = 3.2\,Mpc) of $1.6\pm0.5
\times 10^{38}$ erg\,s$^{-1}$.  Assuming, as usual, a Raymond--Smith
model \citep{Ray77} they derived a plasma
temperature of log($T\rm [K])\,=\,6.8\pm0.3$ and an internal density
of $n_\mathrm{e}\,=\,(0.03\,\pm\,0.01)$\,cm$^{-3}$.  The internal
pressure of $P\approx 4 \times10^5$\,K\,cm$^{-3}$ is again much higher
than the pressure of the ambient warm neutral and ionized medium
($P\approx
10^3-10^4$\,K\,cm$^{-3}$) which led them to suggest that it is
probably this hot gas which is still driving the expansion of the
shell (see e.g., \citet{Wea77}). However, follow--up
observations with \chandra\ indicate that the \rosat source
is more compact. 
In short, only few, possibly pathological holes, are detected in
X--rays.

We should not be too surprised as several authors have pointed out that
the X--ray luminosity of the thermal gas expected to fill the
super-bubbles is well below the \rosat detection limit
\citep{Chu90,Chu95,Mac98,silich}
The expected X--ray luminosity can be
written according to \citet{Mar95} (their Eq. 6) see also
\citet{silich})

\begin{eqnarray}
\label{luminosity}
{\rm L}_{\rm X} &=& 2.12 \times 10^{36}\ {\rm L}_{38}^{33/35} {{\rm
n}_0}^{17/35} {\rm t}_6^{19/35} {\kappa_0}^{4/7}\,{\rm erg\,s}^{-1}\\
\nonumber
&\simeq & 2 \times 10^{36} {\rm L}_{38} \sqrt{{\rm n}_0 {\rm t}_6 {\kappa_0}}\,{\rm erg\,s}^{-1}
\end{eqnarray}

where L$_{38}$ is the mechanical energy expressed in units of
$10^{38}$\, erg\,s$^{-1}$, $n_0$ the ambient density in
atom\,cm$^{-3}$, t$_6$ the age of the super-bubble in Myr and
$\kappa_0$ a dimensionless scaling factor for the classical
conductivity ($\le 1$).
The metalicity $\frac{z}{z_\odot}$ also influences the luminosity in the
X-ray domain. According to \citet{sutherlanddopita} 
the dependence of $L_{\rm X}$ on $z$ can be approximated by 
$L_{\rm X} \propto z^{\frac{1}{3}}$.
In example, in a low metalicity interstellar medium with
$z = 0.1 z_\odot$ the expected X-ray luminosity of an X-ray plasma will only
be reduced by a factor of three, not by an order of magnitude.
In the case of Ho\,II (metalicity: about 0.25$z_{\rm \odot}$, \citet{huntergallagher}) the expected X-ray luminosity will only be marginally reduced.
The ``standard supernova'' described by Eq. \ref{luminosity} will expand into 
an ambient neutral medium, with a typical density ranging between 
$0.1\,\leq n_0\,\leq\,100$.
The age of an expanding structure can be estimated 
from \hi \citep{puche} and FUV data \citep{stewart} to a few tens of 
$10^6$ years. Taking \citet{puche} values of the total mechanical energy needed 
to create the holes of $E \geq 10^{51}\,{\rm erg}$, we can calculate the range 
of expected X-ray luminosity from Eq. \ref{luminosity}. 
Assuming a conductivity of at least $\kappa_0 = 0.5$ gives a range of 
$ 2 \leq \sqrt{{\rm n}_0 {\rm t}_6 {\kappa_0}} \leq 50$.
According to \citep{puche} and \citep{stewart} the mechanical energy ranges
between $0.03 \leq L_{38} \leq 0.3$.
This leads to predicted values of L$_{\rm X} \approx 10^{36}- 10^{37}\,{\rm erg\,s}^{-1}$.
A supernova remnant evolving within a high volume density 
environment has the highest probability to be detected in X-rays with {\it ROSAT\/}.
As an example, source \#7 and \hi hole No. 36 are positional coincident, using $t_6 = 10$ 
\citep{stewart} and $E = 86\times 10^{50}\,{\rm erg}$ \citep{puche} gives 
$L_{38} = 0.27$. With $n_0 = 100\,{\rm cm^{-3}}$ 
and $\kappa_0 = 0.5$ we derive $L_{\rm X} = 4\times 10^{37}\,{\rm erg\,s^{-1}}$.
Given the overall uncertainties this is in good agreement with the observed
value ($9.5\times 10^{37}\,{\rm erg\,s^{-1}}$, see Table \ref{xraysources}).
In a more typical low density environment $n_0 \simeq 0.1\,{\rm cm^{-3}}$ the 
X-ray luminosity is only of order $L_{\rm X} = 1\times 10^{36}\,{\rm erg\,s^{-1}}$, i.e., below the sensitivity of our {\it ROSAT\/} data.

What does this imply? Simply put, the vast majority of super-bubbles
are far too faint to have been detected by {\sl ROSAT}. Those which
have been are in some sense peculiar. One method to boost the X--ray
luminosity, and increase the X--ray emitting lifetime, would be to
invoke mass loading \citep{Arth96}.  Besides, there
remains the possibility that these cases are associated with X--ray
binaries rather than plasma filled super-bubbles.  For example, in the
case of IC\,2574, due to the limited angular resolution of the {\sl
ROSAT} PSPC, it was not possible to differentiate between the X-ray
emission 
originating from hot gas and the thermal emission from an unresolved
individual X-ray source, as now seems likely on the basis of our
\chandra data of IC\,2574.
These unresolved sources do have the same luminosities, of
order $10^{38}$\,erg\,s$^{-1}$, as super-shells. Alternatively, there
could be a contribution to a thermal plasma by supernovae developing
in a dense environment, such as SN1988Z which reportedly reached an
X--ray luminosity of $10^{41}$\,erg\,s$^{-1}$
\citep{fabianterlevich}.

Moreover, it might be difficult to detect coronal gas from expanding
super-bubbles as there could be a conspiracy at work. Perhaps only
(super)bubbles in the making, which are still actively being powered
and hence over-pressured, would be sufficiently bright to be seen by,
for example, {\sl ROSAT}. However, these objects tend to be fairly
small, young, and surrounded by a dense HI shell which would absorb
the X--rays, especially the softer ones. Once a bubble has reached its
final size, after some $10^8$\,yr, the interior has cooled down to
below $10^6$\,K and no X--rays will be detected. It is only with the
new generation of satellites, such as \chandra\ and \xmm\ that we can
hope to settle this issue once and for all.

\section{Summary and conclusions}

We analyzed \rosat PSPC data of the dwarf galaxy Holmberg II. The
\rosat data provide information on faint X--ray emission features
down to a $3\sigma$ unabsorbed limiting flux of $F_\mathrm{X} \simeq 4 \times
10^{-15}$ \fluxunit.  This corresponds to a luminosity threshold of
$L_\mathrm{X} \geq 10^{37}$\lumunit. In total we detected 31
significant X--ray sources located within the \hi\ column density
distribution of Ho\,II.
Using the log($N$) vs. log($S$) relation of \citet{hasinger} we 
expect to detect about 7 XRB sources within the area of Ho\,II \hi gas
disk. 
To avoid confusion with unrelated extragalactic XRB
sources, we investigated further only those X--ray sources which have
counterparts in FUV, \halpha\ or radio continuum emission. This
additional selection criterion is fulfilled by 13 X--ray sources.

In order to determine the nature of faint X--ray sources for which no
high signal--to--noise spectral information can be obtained, we
applied the technique of the hardness ratio 
diagram. For this purpose, we evaluated the intensities in the broad
\rosat 1/4 keV, 3/4 keV and 1.5 keV energy band and calculated the
corresponding X--ray hardness ratios. Depending on the amount of
photoelectric attenuation, an X--ray source will fall within a
specific zone in this diagram. In analogy with the optical regime,
one can consider the variation of the X--ray color due to
photoelectric absorption as the counterpart for reddening of stellar
light by intervening dust.  The X--ray hardness ratio diagram offers a very sensitive tool to constrain the X--ray
emitting process even for faint sources. The faintest source analyzed
in this paper has an X--ray flux of only
$F_\mathrm{X}\mathrm{(0.1-2.4\,keV)}\,=\,11 \times 10^{-15}$
\fluxunit (\#1).

Of the 13 sources detected with confidence and for which counterparts
at other wavelengths have been found, we speculate that of order 3 are
young SNRs, 3 are associated with regions of active star formation, 1
might be the superposed contribution of the coronal X--ray emission of
several hundred O--stars and of order 5 are most likely X--ray
binaries. The strongest source, \#13, was already studied previously by
\citet{zezas}. Whereas they claim a power--law spectrum
and relatively high absorbing column density of about $N_\mathrm{HI}
= 1.2 \times 10^{21}\,\mathrm{cm^{-2}}$, and suggest some source of
accretion on to a compact object, we prefer to interpret this source
as having a softer intrinsic spectrum, and an absorbing column density
of only $N_\mathrm{HI} = 4 \times 10^{20}\,\mathrm{cm^{-2}}$. Our
analysis leads us to propose that this source is most likely a
SNR. This is supported by the analysis published by \citet{miyaji} who
find that some 27\% of the \rosat X--ray emission comes from an
extended source and that the spectrum below 2.5\,keV is best fit by a
thermal plasma. However, their {\it ASCA} observations show that
above 2.5\,keV this source shows a power--law spectrum which they
ascribe to accretion on to an intermediate--mass black hole.

No extended emission which could be attributed to hot gas filling the
\hi\ super-bubbles was detected above our \rosat luminosity limit of
$L_\mathrm{limit}\mathrm{(0.1 - 2.1\,keV)} \geq 10^{37}$\lumunit. This
is not too surprising and in agreement with theoretical estimates
which predict luminosities of L$_{\rm X} \approx 10^{36}-
10^{37}\,{\rm erg\,s}^{-1}$.

As mentioned, observations with \chandra\ and \xmm\ will be needed
and are expected to be especially rewarding for studying the ISM in nearby
(dwarf) galaxies.

\acknowledgements

We like to thank an anonymous referee for careful reading of the
manuscript and a number of good comments which improved the presentation
of the paper.
JK was supported by the Deutsches Zentrum f\"ur Luft-- und Raumfahrt 
under grant No. 50 OR 0103.
FW acknowledges National Sciene Foundation grant AST 96-13717,
EB was supported by a research grant from CONACyT (No.\ 27606--E).

\begin{deluxetable}{lllrrrc}
\tabletypesize{\scriptsize}
\tablecaption{Compilation of \rosat X--ray sources
which coincide in position with counterparts at other
wavelengths.  
\label{xraysources}}
\tablewidth{0pt}
\tablehead{
\colhead{No.} & \colhead{$\alpha_{2000}$} & \colhead{$\delta_{2000}$}  &
\colhead{$F_{\rm X }$} & \colhead{$L_{\rm X }$} & 
\colhead{Source detection} & \colhead{C--band} \\
\colhead{}  & \colhead{[hh:mm:ss]} & \colhead{[$\degr:':''$]} &
\colhead{[$10^{-15}$\fluxunit]} & \colhead{[$10^{37}$\lumunit]} &
\colhead{} & \colhead{source}\\
}
\startdata
1 & 08:19:12 & 70:45:59 & $11.1\pm2.2$  &  $2.4\pm0.6$      & FUV & x\\
2 & 08:18:34 & 70:48:40 & $14.7\pm2.6$  &  $3.3\pm0.6$      & rcs/therm. X--ray & -- \\
3 & 08:18:35 & 70:47:20 & $34.6\pm4.0$  &  $7.7\pm0.9$      & -- & -- \\
4 & 08:18:43 & 70:46:03 & $45.4\pm4.5$  &  $10.1\pm1.0$      & FUV, rcs, X--ray therm. o. powl & -- \\
5 & 08:18:53 & 70:44:45 & $4.8\pm1.5$  &  $1.1\pm0.3$       & FUV, therm. rcs,\halpha\ & x\\
6 & 08:19:03 & 70:43:03 & $4.8\pm1.5$  &  $1.1\pm0.3$       & FUV, therm. rcs, \halpha\  & x\\
7 & 08:19:20 & 70:43:30 & $42.8\pm4.4$  &  $9.5\pm1.0$      & FUV, SNR
rcs, X--ray powl., \halpha\ & x\\
8 & 08:19:17 & 70:42:56 & $140.4\pm7.9$  &  $31.5\pm1.8$    & FUV,
therm. rcs, X--ray therm., \halpha\ & x\\
9 & 08:19:11 & 70:42:01 & $20.1\pm3.0$  &  $4.4\pm0.7$      & FUV,
\halpha\ & x \\
10 & 08:19:00 & 70:41:27 & $16.5\pm2.7$     &  $3.6\pm0.7$   & FUV,
\halpha\ & x\\
11 & 08:19:18 & 70:40:40 & $20.2\pm2.1$     &  $2.2\pm0.4$   & FUV & x \\
12 & 08:19:10 & 70:39:39 & $22.8\pm3.2$     &  $5.1\pm0.7$   & FUV,
rcs,X--ray therm. o. powl & --\\
13 & 08:19:33 & 70:42:18 & $3981.2\pm42.1$  &  $882.1\pm9.4$ & FUV, SNR
rcs, X--ray therm., \halpha\ & x\\
\enddata
\tablecomments{The column labeled ``Source Properties'' lists the most likely
origin for the X--ray flux as well as  information from the complementary 
data. The column labeled ``C-band'' indicates if a source emits significant 
amounts of 1/4 keV radiation, suggesting that the X--ray source is located in
front of the gaseous body of Ho\,II. The abbreviation ``rcs'' denotes radio 
continuum source. The radio data is published by \citet{tonguewestpfahl}, the 
optical data by \citet{rhode} and the FUV data by \citet{stewart}.}
\end{deluxetable}




\end{document}